\documentclass[reprint, superscriptaddress, secnumarabic, amssymb, nobibnotes, aps, prl]{revtex4-1}

\usepackage{graphicx}
\usepackage{epstopdf}
\usepackage[T1]{fontenc}
\usepackage[latin9]{inputenc}
\usepackage{amsbsy}
\usepackage{gensymb}
\usepackage[T1]{fontenc}
\usepackage[latin9]{inputenc}
\usepackage{amsmath}
\usepackage{amssymb}
\usepackage{bbm}
\usepackage{braket}
\usepackage{xcolor}
\allowdisplaybreaks
\usepackage{graphicx}
\usepackage[colorlinks=true]{hyperref}  
\hypersetup{
    bookmarks=true,         
    unicode=false,          
    pdftoolbar=true,        
    pdfmenubar=true,        
    pdffitwindow=false,     
    pdfstartview={FitH},    
    pdftitle={Muon spin spectroscopy study of the noncentrosymmetric superconductor TaOs},    
    pdfauthor={D. Singh, Sajilesh K.P, A. D. Hillier, R. P. Singh},    
    pdfsubject={},   
    pdfcreator={},   
    pdfproducer={}, 
    pdfkeywords={} {} {}, 
    pdfnewwindow=true,      
    colorlinks=true,       
    linkcolor=blue, 
    citecolor=blue,        
    filecolor=magenta,      
    urlcolor=blue           
} 
\usepackage[normalem]{ulem}

\newcommand{\figref}[1]{Fig.~\ref{#1}}

\begin{document}
\title{\textrm{Superconducting properties of the non-centrosymmetric Superconductors TaXSi (X= Re, Ru) }}
\author{Sajilesh~K.~P.}
\affiliation{Department of Physics, Indian Institute of Science Education and Research Bhopal, Bhopal, 462066, India}
\author{R.~P.~Singh}
\affiliation{Department of Physics, Indian Institute of Science Education and Research Bhopal, Bhopal, 462066, India}
\date{\today}
\begin{abstract}
\begin{flushleft}
\end{flushleft}
We have investigated the ternary noncentrosymmetric superconductors TaXSi (X=Re, Ru) by magnetization, resistivity, and specific heat measurements. The samples crystallize in orthorhombic TiFeSi structure having superconducting transition T$_{c}$ = 5.32 K and 3.91 K, for TaReSi and TaRuSi respectively. Specific heat measurements indicated an s-wave nature of both materials with a moderately coupled nature. However, a low value of specific heat jump and the concave nature of the upper critical field suggests a nontrivial superconducting gap.

\end{abstract}

\maketitle
\section{Introduction}
Superconductivity in noncentrosymmetric (NCS) systems has sparkled a renewed research interest owing to their fascinating properties of fundamental interest, such as anisotropic superconducting gap, time reversal symmetry breaking, and the presence of Majorana quasiparticles \cite{Bauer2004,topo1,topo2}. Systems with strong spin-orbit coupling have been the prime candidate to show nontrivial band topology, leading to topologically protected zero-energy surface modes \cite{Hsoc1,Hsoc3,topo3}. NCS materials are remarkable in this regard with its intrinsic Rashba-type antisymmetric spin-orbit (ASOC) interactions that lifts the spin degeneracy of the electronic bands at the Fermi level and generate complex spin textures. \cite{rashba1,rashba2,rashba3,EBA}. This, in general, can lead to Cooper pair of mixed singlet-triplet character, leading to a broken time reversal symmetry and anisotropy in the superconducting gap \cite{vm,kv,fujimoto,nmr,Bauer2004}. Akin to topological insulators, this nontrivial pairing results in various types of protected zero energy states at the edge or surface of NCS materials \cite{edg1}. Furthermore, topologically protected zero-energy boundary modes also occur in NCSs with an anisotropic superconducting gap \cite{topo2,topo4,TOS}.   
\\  
\\
The current research in noncentrosymmetric materials is focused on finding new materials with high ASOC and establishing a relation between the strength of ASOC and its influence on the superconducting ground state. The discovery of CePt$ _{3} $Si \cite{Bauer2004} with nodes in the superconducting gap has revived interest in this field. Experimental evidence suggests a strong ASOC (50-200 meV) \cite{kv} in this material has triggered the presence of line nodes. Another remarkable evidence of ASOC dependence on the gap structure was visible when Pd was replaced with Pt in Li$ _{2} $(Pd,Pt)$ _{3} $B \cite{LPB1,LPB2}. This material has shown the presence of triplet along with singlet, which challenges the primary concepts explaining the superconducting phenomenon. Several other members of the noncentrosymmetric family have also shown anisotropic \cite{MAC,CPS,LPB3,KCA,CIS1} gap structure while only a handful of compounds has shown time reversal symmetry breaking \cite{LNC1,LNC2,LI,RZ3,RH2,RT,CPA,LR}. Despite evidence supporting the ASOC dependence on the superconducting ground state, several materials have shown conventional isotropic BCS superconductivity \cite{LPS,LRS,LS2,LS3,RTa,CIS,LMP}. Among which, LaPt$_{3} $Si with a similar structure as CePt$ _{3} $Si and very strong ASOC has failed to show any unconventional behavior \cite{LP3S}. At the same time, few noncentrosymmetric materials with very low ASOC have also shown triplet presence, and nodal superconductivity \cite{LNC3}. It is also suggested for CePt$ _{3} $Si that the ferromagnetic ordering might have caused the nodal behavior, which is absent for the case of LaPt$ _{3} $Si. It raises questions on the selective observation of unconventional superconductivity in NCS systems and their explicit dependence on the strength of ASOC.
\\

\begin{figure*}[t] 
	\includegraphics[width=2.0\columnwidth, origin=b]{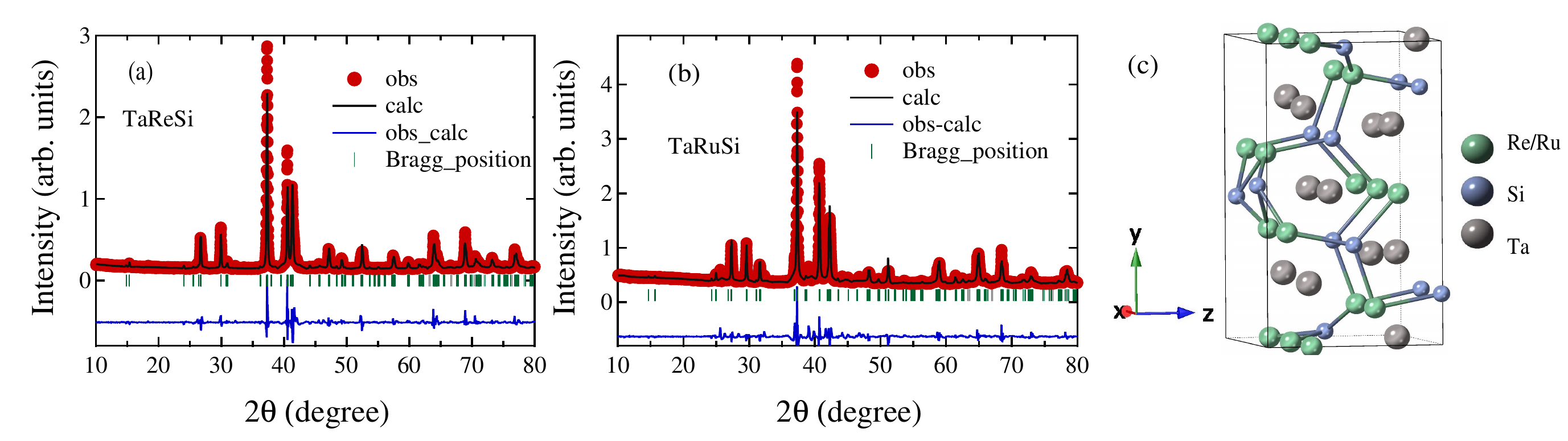}
	\caption{Powder x-ray diffraction pattern for (a) TaReSi and (b) TaRuSi sample obtained at room temperature using Cu K$ _{\alpha} $ radiation (red line). The solid black line shows the Rietveld refinement whereas the blue line shows the difference between the observed and calculated one. (c) Crystal structure of TaXSi samples. }
	\label{fig1}
\end{figure*}

Ternary noncentrosymmetric materials give an excellent platform to investigate the role of ASOC on the superconducting ground state similar to the case of Li$ _{2} $(Pd,Pt)$ _{3} $B. It is easy to play with the strength of ASOC, which can be tuned for the case of ternary materials by replacing the constituent elements. For the present study, we have selected TaXSi, where X represents Re/Ru. Both the materials crystallize into a noncentrosymmetric orthorhombic TiFeSi-type structure (space group $ I m a 2 $). The TiFeSi-type structure is a superstructure modification of ordered hexagonal Fe$_{2}$P structure. This structural transition occurs due to small displacements of atoms from their ideal hexagonal position, with a reduction in symmetry from hexagonal to body centered orthorhombic structure. The superconducting properties of ternary equiatomic systems are strongly influenced by the crystal structure. Among these, ZrRuP with hexagonal Fe$_{2}$P structure has shown the highest T$_{c}$ at 13 K, while the TiFeSi family, in general, has shown low T$_{c}$. The high T$_{c}$ in hexagonal ZrRuP is expected to originate from the strong electron-phonon interaction, where the electron-phonon coupling constant has a value of 1.25 \cite{ZrRuP}. Furthermore, the initial band structure calculation revealed the enforced semi-metal nature of TaRuSi with possible topological nature. Hence, it will be interesting to look for the implications of the nature of the band structure on the superconducting ground state \cite{band1,band2}. Also, the TiFeSi-type structure falls under the globally stable nonsymmorphic symmetry, which is favorable for topological material \cite{nonsymo}.   Superconducting transition in TaXSi materials was reported in 1985, while the nature of the superconductivity and normal state property remained unexplored \cite{GV}. Re, being a heavy transition element, can induce a strong ASOC in TaReSi, compared to TaRuSi where Ru is a comparatively light element. This difference is expected to have effects on the superconducting properties as well as the ground state. In this paper, we have studied the superconducting as well as the normal state properties of both the samples using resistivity, specific heat, and magnetization measurements. A small jump in specific heat along with the concave nature of the upper critical field observed for both samples might indicate an unconventional gap feature.\\

\section{Experimental Details}
Polycrystalline TaXSi (X = Re, Ru) samples were prepared using a standard arc melting technique. High purity Ta (99.99\% ), Re (99.99\%),(or Ru (99.99\% )), and Si (99.99\%) were taken in a stoichiometric ratio and melted on a water-cooled copper hearth under high purity Argon gas. For better phase purity Ta and X were melted together at the first step, which then melted with Si. This method reduces the weight loss in the melting process. The resulting ingot formed with the negligible mass loss was flipped and remelted several times to improve the homogeneity. Phase purity and crystal structure of the sample was confirmed by room temperature x-ray diffraction measurement using a PANalytical diffractometer equipped with CuK$\alpha$ radiation($\lambda$ = 1.5406 \AA). Magnetization measurements were done using a superconducting quantum interference device (MPMS 3, Quantum Design) at various temperatures and field ranges. The electrical resistivity and heat capacity measurements of the sample were performed using a Physical Property Measurement System (PPMS, Quantum Design).

\begin{figure} 
	\includegraphics[width=1.0\columnwidth, origin=b]{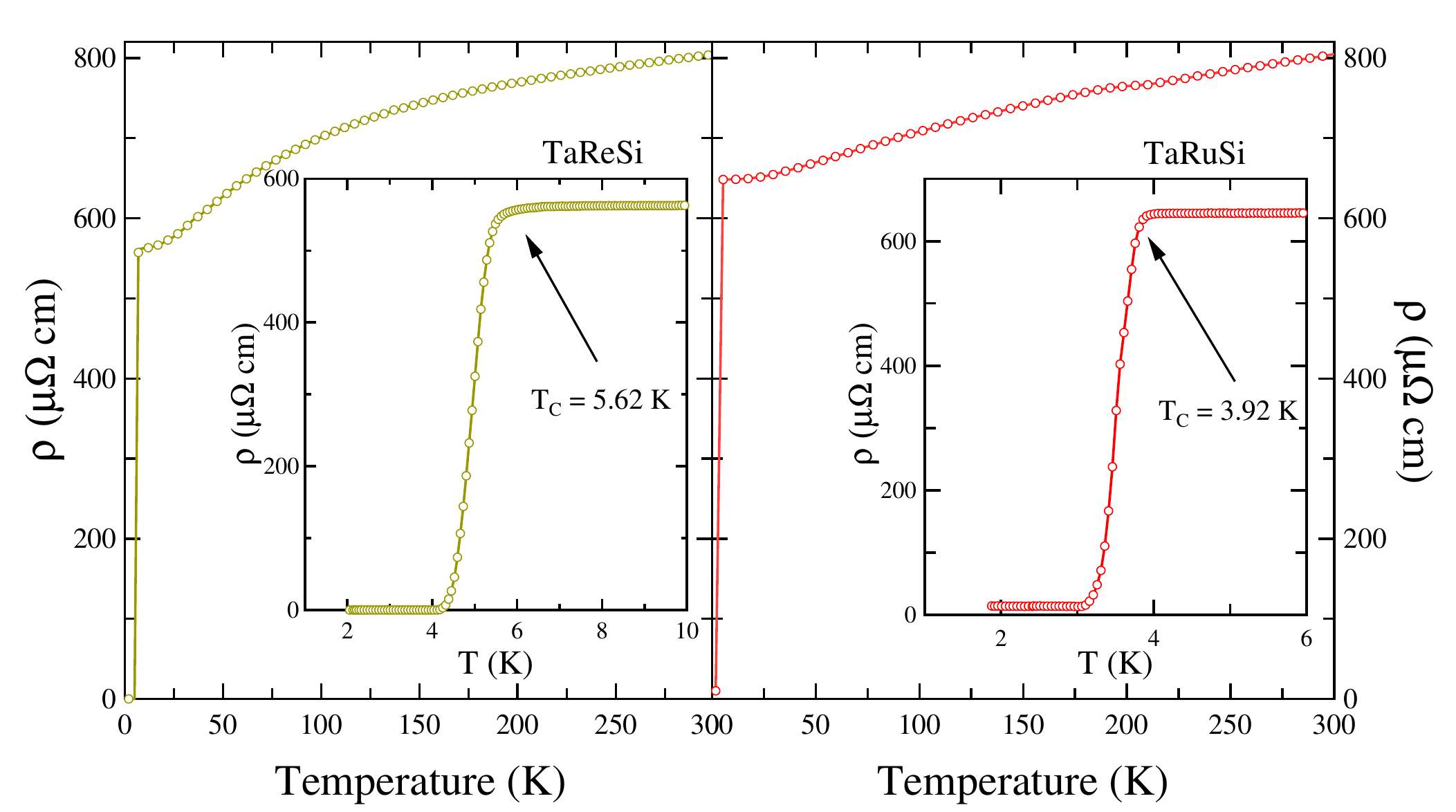}
	\caption{Temperature dependence of resistivity in the range 1.8 K $ \leq $ T $ \leq $ 300 K. The inset shows the drop in resistivity at the superconducting transition, T$ _{c} $ = 5.62 $\pm$ 0.05 K and 3.92 $\pm$ 0.05 K respectively for TaReSi and TaRuSi. }
	\label{fig2 }
	\label{fig2}
\end{figure}

\begin{figure*} 
 \includegraphics[width=2.0\columnwidth, origin=b]{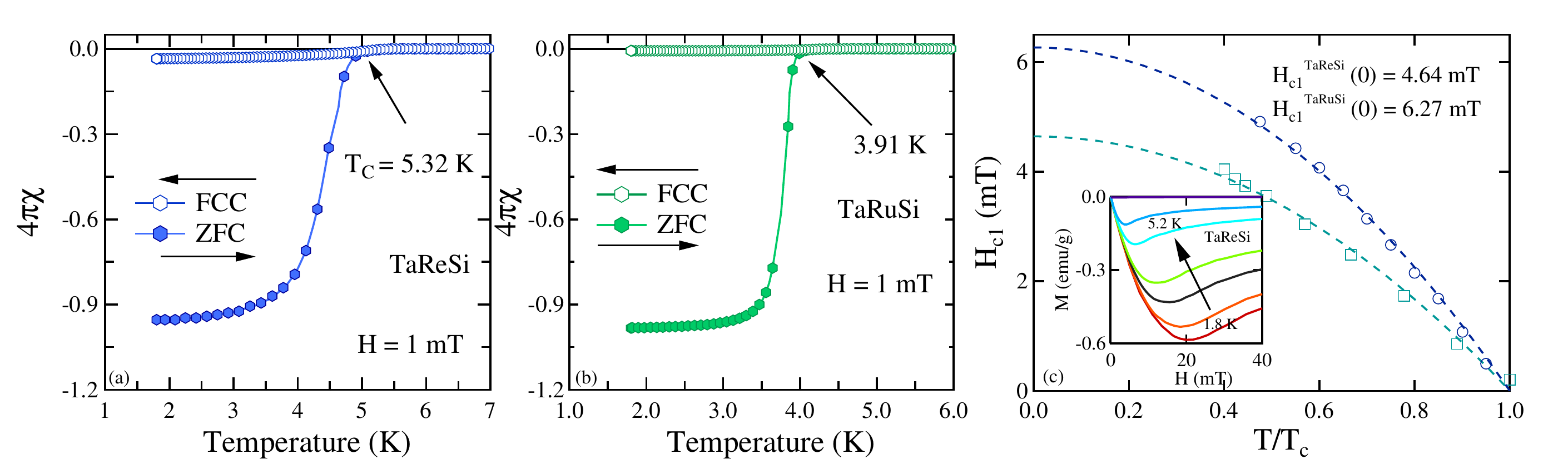}
 \caption{ (a) and (b) Temperature dependence of magnetic moment collected via zero field cooled warming (ZFC), and field cooled cooling (FCC) methods under an applied field of 1 mT. Onset of diamagnetic signal was observed at T$ _{C}^{onset} $ = 5.32 K and 3.91 K for TaReSi and TaRuSi respectively. (c) Lower critical field estimated from M-H curve using the G-L equation.  }
 \label{fig3}
 \end{figure*}

\section{Results and Discussion}
\subsection{Sample characterization}
\figref{fig1} shows the X-ray diffraction pattern for the TaXSi samples, collected at ambient pressure and temperature. The samples crystallize into noncentrosymmetric orthorhombic FeSiTi structure type (Space group $ I m a 2 $). The obtained diffraction pattern fits very well with the reported data showing the phase purity of the samples. \figref{fig1} (c) shows the body-centered orthorhombic structure of TaXSi. The fitted lattice cell parameters for both the compounds are enlisted in Table \ref{xrd table}.

\begin{table}[h!]
	\caption{Crystallographic parameters of TaXSi}
	\label{xrd table}
	Crystal structure: Orthorhombic TiFeSi type\\
	Space group: $ I m a 2 $ (46)
	\begin{center}
		\begin{tabular*}{1.0\columnwidth}{l@{\extracolsep{\fill}}lll}\hline\hline
		Parameters& TaReSi& TaRuSi\\
		\hline
		\\[0.5ex]                                  
		a&  6.972(7)& 7.132(4)\\
		b& 11.574(1)& 11.292(2)\\
		c& 6.657(6)& 6.547(6)\\
		&$ \alpha $ = $ \beta $ = $ \gamma $ = 90$ \textdegree $
		\\[0.5ex]
		\hline\hline
		\end{tabular*}
		\par\medskip\footnotesize
	\end{center}
\end{table}

\subsection{Resistivity}
 The \figref{fig2} displays the temperature dependence of resistivity for TaReSi and TaRuSi in the range 1.8 K $ \leq $ T $ \leq $ 300 K. Both the samples showed a decrease in resistivity as temperature decreases, showing metallic behavior in the whole temperature range. A drop in resistivity was observed at T$ _{C}^{onset} $ = 5.62 $\pm$ 0.05 K and 3.92 $\pm$ 0.05 K, respectively, suggesting the onset of superconductivity. The residual resistivity ratio is low, indicating resistive property in the normal state region is sensitive to defects and other scattering centers present in the polycrystalline samples.
 
 \subsection{Magnetization}
 A temperature-dependent DC magnetic susceptibility measurement has shown a marked drop at around T$ _{c}^{onset} $ = 5.32 K and 3.91 K at 1 mT (Fig. \ref{fig3}), suggesting the occurrence of superconductivity for Re and Ru variant, respectively. A type-II nature of both the samples is visible from the flux pinning nature during FCC measurement. We have used the sample in rectangular cuboid shape for magnetization measurement, and the superconducting volume fraction is close to 100\% corresponding to a full diamagnetic shielding. This rules out any significant impurity in both samples. A low field magnetization was carried out to estimate the lower critical field H$ _{c1} $ for the compounds. The deviation from linear behavior in the magnetization curve is taken as the H$ _{c1} $ at that particular temperature. An extrapolation of H$ _{c1} $ (T) using G-L equation H$ _{c1} $(T) = H$ _{c1} $(0)(1-t$^{2}$), where t = T/T$ _{c} $ gives H$ _{c1} $(0) = 4.64 $ \pm $ 0.08 mT and 6.27 $ \pm $ 0.04 mT for TaReSi and TaRuSi respectively.

 \subsection{Upper critical field}
\begin{figure} 
	\includegraphics[width=1.0\columnwidth, origin=b]{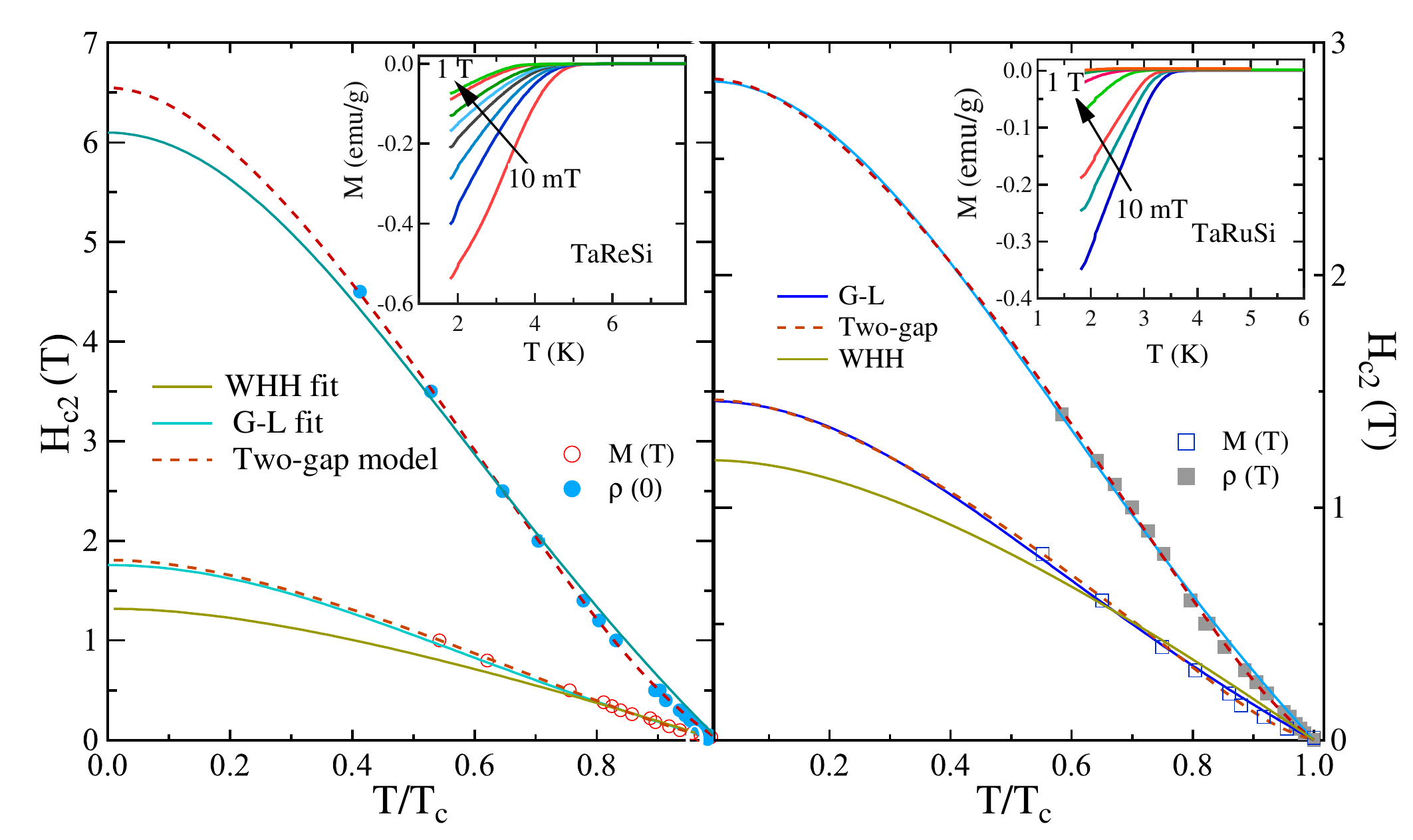}
	\caption{ Estimation of the upper critical field by magnetization and resistivity measurements. The H$ _{c2} $(T) determined from magnetization is fitted by WHH, G-L, and two-gap model. The WHH model has failed to trace the data points, while the G-L and the two-gap model has successfully estimated the H$ _{c2} $(T). The H$ _{c2} $(T) determined from resistivity measurements has shown comparatively high value, probably due to surface effects. The resistivity data is fitted using the two-gap and G-L model, as shown.  }
	\label{fig4}
\end{figure}

 The upper critical field H$ _{c2} $ for the compounds is determined by both magnetization as well as resistivity measurements in the field range 10 mT $ \leq $ H $ \leq $ 1 T. The transition temperature was seen shifting towards lower temperatures as the field increases, with transition becoming broader. The onset of superconductivity in magnetization/resistivity at each field is taken as the value of H$ _{c2} $. The H$ _{c2} $ curve obtained from magnetization data in the 0 - T$ _{c} $ range can be extrapolated using the WHH model, considering the effects of orbital breaking, Pauli spin paramagnetism ($ \alpha $), and spin-orbit scattering parameter ($ \lambda_{so} $) \cite{WHH1,WHH2}. According to this model, H$ _{c2} $ can be implicitly explained by the expression,

 \begin{multline}
   ln\left(\frac{1}{t}\right) = \left(\frac{1}{2} + \frac{i\lambda_{so}}{4\gamma}\right)\psi\left(\frac{1}{2} + \frac{h+\frac{1}{2}\lambda_{so}+i\gamma}{2t}\right)+\\  
  \left(\frac{1}{2} - \frac{i\lambda_{so}}{4\gamma}\right)\psi\left(\frac{1}{2} + \frac{h+\frac{1}{2}\lambda_{so}-i\gamma}{2t}\right)-\psi\left(\frac{1}{2}\right)
\label{whh}
\end{multline} 

where t=$ \frac{T}{T_{c}} $ is the reduced temperature. $ \lambda_{so} $ is the spin-orbit scattering parameter, $ \psi $ is the digamma function, $ \gamma $ = $ \sqrt{(\alpha_{M}h)^{2}-(\frac{1}{2}\lambda_{so})^{2}} $, $ \alpha_{M} $ is the Maki parameter, and $h$ is the dimensionless form of the upper critical field given by $h$ = $ (4/\pi^{2})(H_{c2}/\mid dH_{c2}/dT\mid_{T_{c}}) $. Extrapolating temperature dependence of H$ _{c2} $ for the two samples with  $ \alpha_{M} $ = 0.19, 0.24  and $ \lambda_{so} $ = 0, 0 respectively for TaReSi and TaRuSi gave a best fit using the model and is shown in \figref{fig4}. The upper critical field using WHH model can be approximated by

\begin{equation}
H_{c2}^{orbital}(0) = -0.693\times T_{c}\times\left.\frac{dH_{c2}(T)}{dT}\right|_{T=T_{c}}
\label{eqn4:whh}
\end{equation}

where $ \alpha $ = 0.528$ \frac{dH_{c2}(T)}{dT}|_{T=T_{c}} $. Combining the expressions, we get $ H_{c2} $(0) = 1.35 T for TaReSi and 1.25 T by WHH model. However, this model is insufficient to reproduce the data points due to concave upward nature H$ _{c2} $ for both samples, prominent for TaRuSi, giving an underestimated value of H$ _{c2} $(0) . This can be arise from various reasons such as   localization effects \cite{hc21}, twisting of electron orbits by a magnetic field \cite{hc22}, dimensional cross over \cite{hc23}, multi-gap behavior etc \cite{hc24}.

The slight upturn nature of H$ _{c2} $(T), prominent for the case of TaRuSi, is similar to the case reported for the two-gap superconductors MgB$_{ 2} $, YNi$ _{2} $B$ _{2} $C, LuNi$ _{2} $B$ _{2} $C, 2$H$-NbSe$_{2}$ \cite{MgB2,YNBC,YNBC2,NbSe2} . Hence, we have attempted to describe the H$_{c2}$(T) curve using the two-gap model, according to which, H$_{c2}$(T) is described by the parametric equation,  

 \begin{multline}
ln\left(\frac{1}{t}\right) = \left[U(s) + U(\eta s) + \frac{\lambda_{0}}{w}\right] + \\  \left( \frac{1}{4} \left[U(s) - U(\eta s) - \frac{\lambda_{-}}{w}\right]^{2} +  \frac{\lambda _{eh}\lambda _{he}}{w^{2}} \right) ^{1/2}\\
H_{c2}   = \frac{ 2\phi_{0}Ts}{D_{e} } \;\;\;\;\;\;  \eta = \frac{D_{h}}{D_{e}}\\
U(s) = \psi(s+ 1/2) - \psi(1/2)\;\;\;\;\;\;\;\;\;\;\;\;\;\;
\label{whh}
\end{multline} 

Here, $ \lambda_{-} $ = $ \lambda_{ee} - \lambda_{hh} $, $ \lambda_{0} = (\lambda_{-}^{2} + 4\lambda_{eh}\lambda_{he})$, $ w = \lambda_{ee}\lambda_{hh} - \lambda_{he}\lambda_{eh} $. The variables, $ \lambda_{ee}, \lambda_{hh}, \lambda_{eh}, \lambda_{he} $ are the matrix elements of the BCS coupling constants. $ D_{e} $ and $ D_{h} $ are the electron and hole diffusivity. $ \phi_{0} $ is the flux quantum and $ \psi (s) $ is the digamma function. Though the fitting seems to be in good agreement with the experimental data, we must admit that there remains a questionable reliability of the fitting parameters since the fit was done for a large number of parameters. However, extrapolating to zero temperature yields the values of H$ _{c2} $(0) as 1.81 T and 1.46 T respectively for TaReSi and TaRuSi, close to that obtained from G-L fitting.

According to Maki theory \cite{maki}, the the upper critical field at 0 K is related to $ \alpha $ by  the relation, H$ _{c2} $(0) = $ \alpha $H$ _{P} $(0)/$ \sqrt{2} $ where H$ _{P} $ is the zero temperature Pauli limiting field. H$ _{P} $(0) can be relate to H$ _{P}^{BCS} $, the BCS value for paramagnetic limiting field by the equation H$ _{P} $(0) = H$ _{P}^{BCS} \sqrt{1+\lambda_{e-ph}}$. Substituting $ \alpha $ = 0.18, 0.24 and $ \lambda_{e-ph} $ = 0.63 and 0.58 for TaReSi and TaRuSi respectively, we get H$ _{c2} $(0) = 1.63 T and 1.52 T. This value is in close agreement with prediction from the G-L formula which describe the temperature dependence of H$ _{c2} $ as 
 \begin{equation}
H_{c2}(T) = H_{c2}(0)\left[\frac{(1-t^{2})}{(1+t^2)}\right]. 
\label{hc2}
\end{equation}

A fitting employed using this relation gave H$_{c2} $(0) = 1.76 $ \pm $ 0.03 T and 1.46 $ \pm $ 0.02 T respectively for TaReSi and TaRuSi. The coherence length is calculated to be 137 $ \pm $ 2 \text{\AA} and 114 $ \pm $ 2 \text{\AA} respectively for Re and Ru variant using $ \xi_{GL} $ = ($\phi_{0}$/2$ \pi H _{c2}(0) $)$ ^{1/2} $ ( $\phi_{0} $ = 2.07 $ \times $ 10$ ^{-15} $Tm$ ^{2} $) and the magnetic penetration depth for the sample $ \lambda_{GL}(0) $ is estimated using the relation 
\begin{equation}
H_{c1}(0) = \frac{\Phi_{0}}{4\pi\lambda_{GL}^2(0)}\left(\mathrm{ln}\frac{\lambda_{GL}(0)}{\xi_{GL}(0)}+0.12\right)   
\label{eqn6:ld}
\end{equation} 

which is obtained as 3373 $ \pm $ 87 \text{\AA} and 2766 $ \pm $ 62 \text{\AA}. Following the penetration and coherence length, the Ginzburg- Landau parameter for the samples can be found out as 25 $ \pm $ 1 and 18 $ \pm $ 1.

However, the temperature dependence of H$_{c2}$ determined from resistivity measurements has shown a relatively high value. Such a high value can be arisen due to surface or filamentary effects. Here, much stronger scattering of electrons at grain boundaries can reduce the mean free path, which in turn reduces the coherence length, increasing the upper critical field. Also, a higher residual resistivity value ($\rho_{0}$ = 559 and 647 $\mu \ohm$ cm for TaReSi and TaRuSi respectively) indicates higher density of defects/disorder in the system. The magnetic flux line, in this case, can pin to these defects and hence reducing the effects of the orbital pair breaking, increasing the upper critical field. Similar high upper critical field is reported for LaPtSi, BaPtSi$_{3}$, LaIrSi$_{3}$ \cite{LPS, BaPtSi, LS3}. We have extrapolated the data using both the G-L model and the two-gap model as done for the magnetization data. Similar to magnetization data, we have obtained better fitting of the data points using the two-gap model, giving rise to H$_{c2}$(0) = 6.55 T and 2.84 T respectively for TaReSi and TaRuSi.

 \begin{figure} 
	\includegraphics[width=1.0\columnwidth, origin=b]{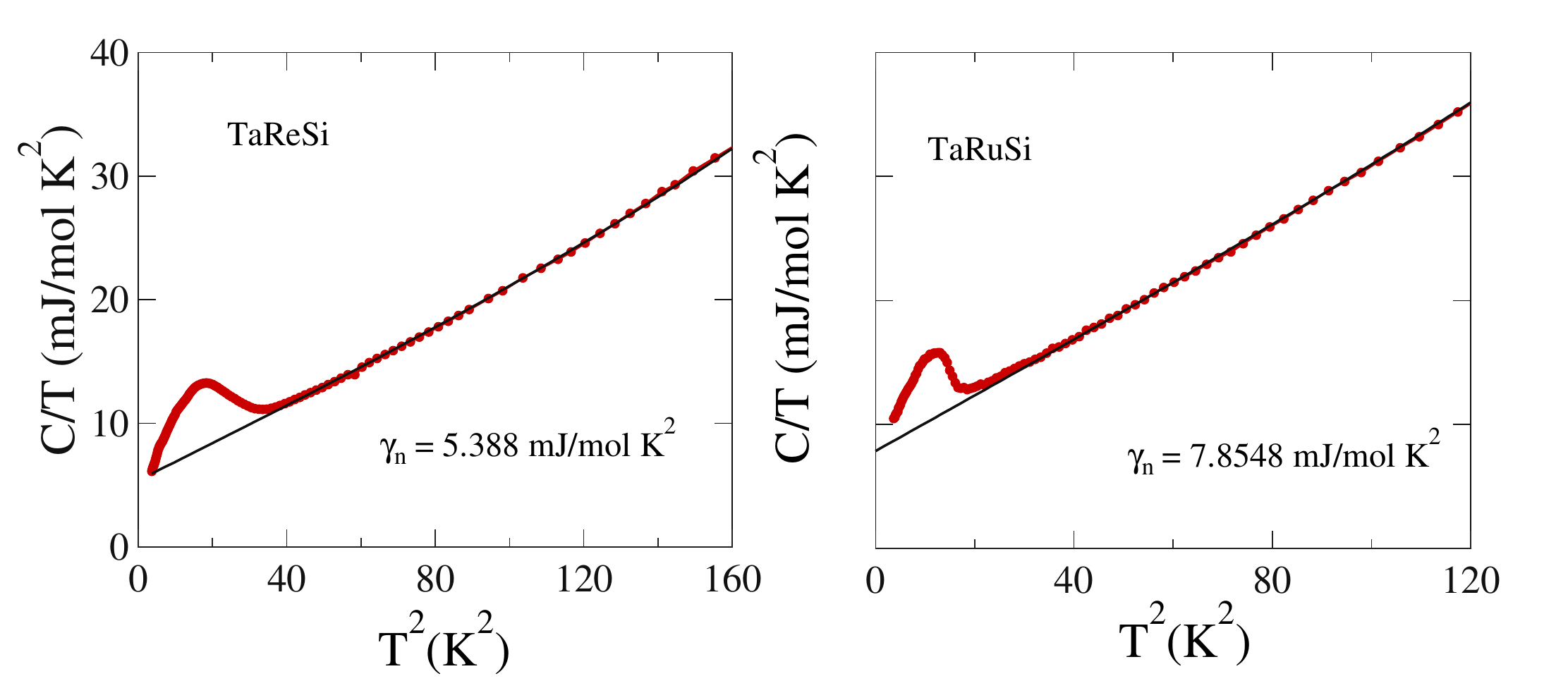}
	\caption{C/T vs T$ ^{2} $ for the samples shows a specific heat anomaly at T$ _{c} $ = 5.4 and 3.86 K respectively for TaReSi and TaRuSi. The black line shows the fit to the curve using Eq. \ref{eqn3:hc}}
	\label{fig5}
\end{figure}

\subsection{Specific Heat}
Heat capacity measurement was carried out in the temperature range 1.9 K $ \leqslant $ T $ \leqslant $ 15 K showing the bulk nature of superconductivity in both compounds with a jump at T$ _{c}^{mid} $ = 5.4 $ \pm $ 0.1 K and 3.86 $ \pm $ 0.1 K respectively for TaReSi and TaRuSi. The contribution to specific heat in the normal state at low temperatures arises from electronic as well as the phononic origin, which can be well explained by the relation
\\
\begin{equation}  
\frac{C}{T}=\gamma_{n}+\beta_{3} T^{2} + \beta_{5}T^{4} . 
\label{eqn3:hc}    
\end{equation}
\\
Here $ \gamma_{n} $ accounts for the electronic contribution while $ \beta_{3} $ and $ \beta_{5} $ represents the phononic contribution. Figure. \ref{fig5} shows the specific heat data plotted as C/T vs T$^{2}$ for both samples. Fitting yields the values as $ \gamma_{n}$ = 5.38 $\pm$ 0.07  mJ/mol K$^{2}$, $\beta_{3}$ = 0.151 $\pm$ 0.001  mJ/mol K$^{4}$ and $ \beta_{5} $ = (6.774 $ \pm $ 0.001)$ \times $ 10$ ^{-7} $mJ/mol K$^{6}$ for TaReSi while for TaRuSi it is $ \gamma_{n}$ = 7.85 $\pm$ 0.03  mJ/mol K$^{2}$, $\beta_{3}$ = 0.224 $\pm$ 0.001  mJ/mol K$^{4}$ and (7.112 $ \pm $ 0.001)$ \times $ 10$ ^{-8} $mJ/mol K$^{6}$.\\

 A number of parameters characterizing the compounds can be extracted from the fitted parameters.  Using the value of $\beta_{3}$, Debye temperature $ \theta_{D} $ of the sample can be estimated using the relation
\begin{equation} 
\theta_{D}= \left(\frac{12\pi^{4}RN}{5\beta_{3}}\right)^{\frac{1}{3}}
\label{eqn4:dt}  
\end{equation}

\begin{figure} 
      \includegraphics[width=1.0\columnwidth, origin=b]{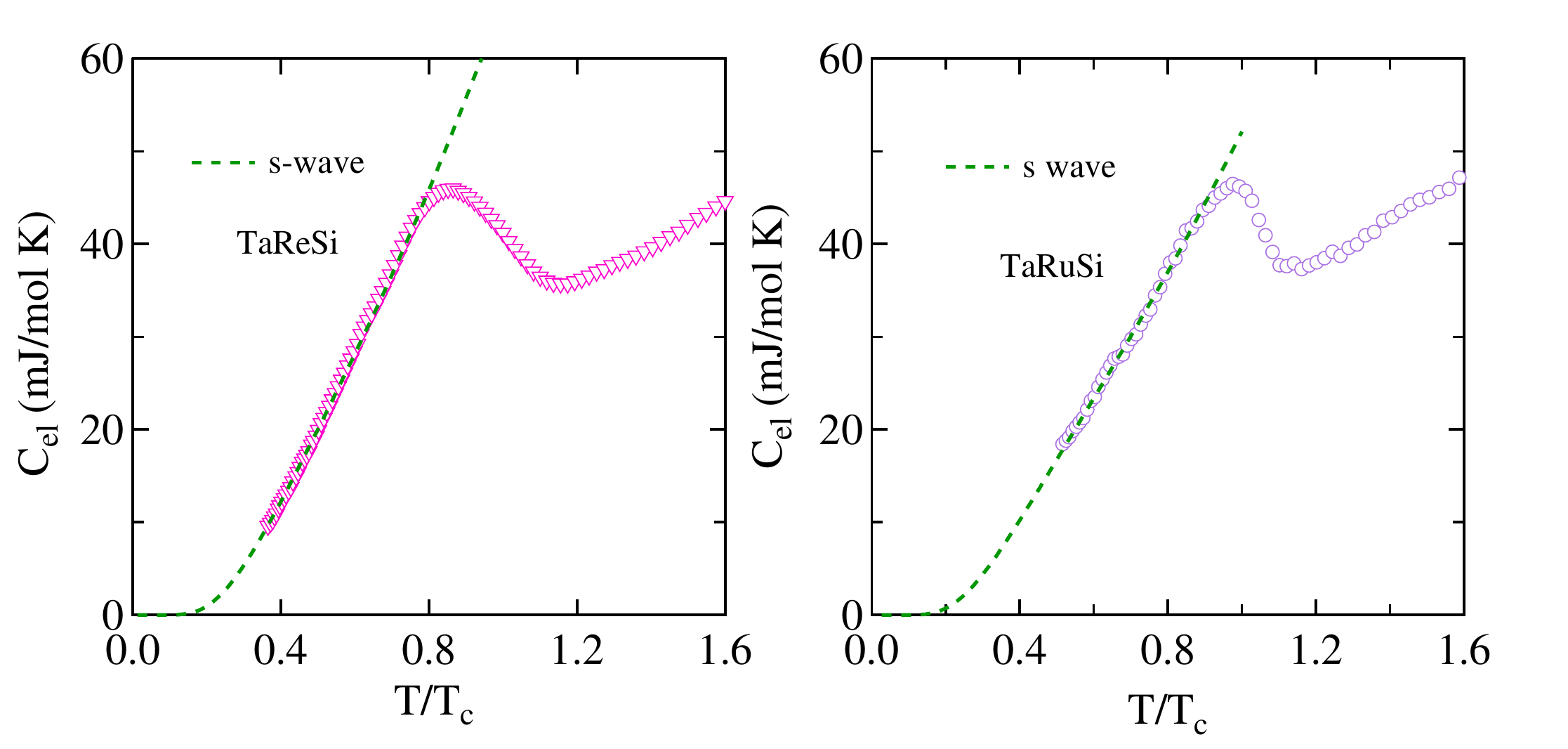}
      \caption{A temperature dependence of electronic specific heat till 1.9 K shows an s-wave behavior (dotted line) with isotropic gap value $\frac{\Delta(0)}{k_{B} T_{c}} $ = 1.4 $ \pm $ 0.04  and 1.36 $ \pm $ 0.04 respectively.}
      \label{fig6}
      \end{figure}

Substituting R, the universal gas constant and N = 3, the number of atoms per formula unit, we have obtained $ \theta_{D} $ = 338 $ \pm $ 2 K and 296 $ \pm $ 2 K respectively for TaReSi and TaRuSi.  An estimation of $ \lambda $, the electron-phonon scattering parameter, which quantizes the strength of electron-phonon coupling can be calculated by McMillans formulae \cite{mac}
\begin{equation}
\lambda_{e-ph} = \frac{1.04+\mu^{*}ln(\theta_{D}/1.45T_{c})}{(1-0.62\mu^{*})ln(\theta_{D}/1.45T_{c})-1.04 }
\label{eqn8}
\end{equation} 

where $ \mu^{*} $ is 0.13 for intermetallic superconductors. Inserting the value of Debye temperature $ \theta_{D} $, the obtained values are 0.63 $ \pm $ 0.02 and 0.58 $ \pm $ 0.02 respectively for TaReSi and TaRuSi. This places both the superconductors in the moderately coupled family. The value of the Sommerfeld coefficient can be inserted in the equation
\begin{equation} 
\gamma_{n}= \left(\frac{\pi^{2}k_{B}^{2}}{3}\right)D_{C}(E_{\mathrm{F}})
\label{eqn5:ds}  
\end{equation}
to determine the density of states at the Fermi surface $ D_{C}(E_{\mathrm{F}}) $. This gives 2.28 $ \pm $ 0.03 and 3.34 $ \pm $ 0.01 $ \frac{states}{eV f.u} $ for TaReSi and TaRuSi respectively. The electronic contribution to specific heat can be calculated by directly subtracting the phonic contribution, using the relation C$ _{el} $ = C $ - $ $ \beta $T$^{3}  $ $ - $ $ \beta $T$^{5}  $. Figure. \ref{fig6} shows the electronic specific heat, C$ _{el} $ plotted against normalized temperature. A normalised jump in electronic specific heat, $ \frac{\Delta C_{el}}{\gamma_{n} T_{c}} $ for both the samples are close to 1,(1.07 and 0.91 for TaReSi and TaRuSi respectivly) which is less than the BCS approximation. An isotropic s-wave model \cite{LPG} in the dirty limit regime can be used to trace the data points in the superconducting region, giving a normalized superconducting gap as  $ \frac{\Delta(0)}{k_{B} T_{c}} $ = 1.4 $ \pm $ 0.04  and 1.36 $ \pm $ 0.04.

      A set of four equations as explained in Refs. \cite{RZ2,NbOs} is simultaneously solved to get BCS coherence length $ \xi_{0} $, mean free path $ l $, Fermi velocity V$ _{\textit{f}} $, superconducting carrier density $n$, effective mass $m ^{*} $. The obtained ratio $ \xi_{0} $/ $l$ = 9.66 and 11.75 for TaReSi and TaRuSi indicate the dirty limit nature of the samples. The calculated parameters are tabulated in Table \ref{elec propr}. Using the value of $n$, the Fermi temperature for the system can be extracted from the relation,

\begin{equation}
 k_{B}T_{F} = \frac{\hbar^{2}}{2}(3\pi^{2})^{2/3}\frac{n^{2/3}}{m^{*}}, 
\label{eqn13:tf}
\end{equation}   

\begin{figure}
	\includegraphics[width=1.0\columnwidth]{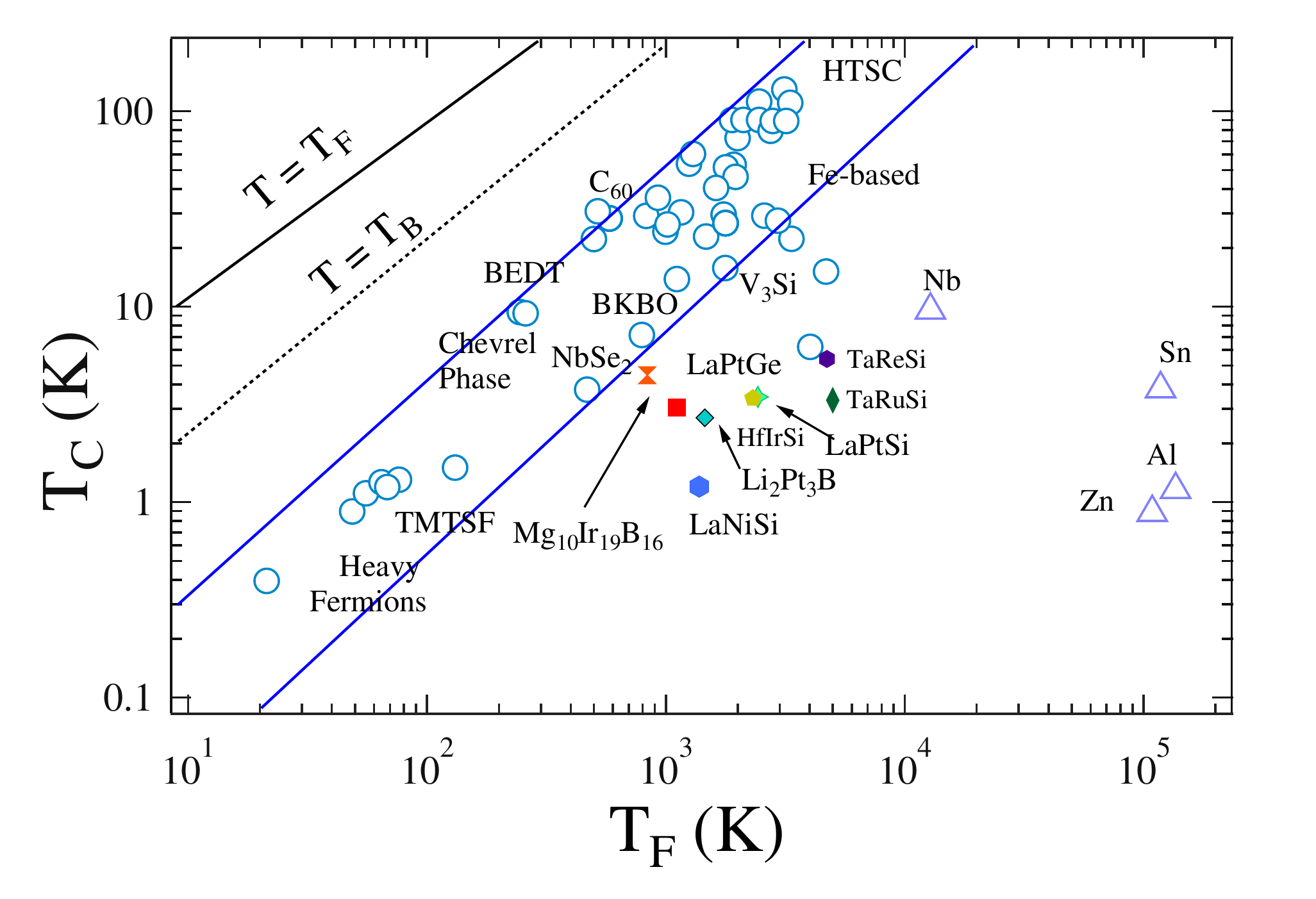}
	\caption{\label{Uemura}Uemura plot showing  $T_{c} $ Vs. the effective Fermi temperature $T _{F} $. The blue band represents different families of superconductors with unconventional properties. The dashed line corresponds to the Bose-Einstein condensation temperature. The positions of TaReSi and TaRuSi indicate a conventional nature of these materials.}
\end{figure}

This gave the Fermi temperatures for the system as T$ _{F} $ = 4997 K and 5066 K respectively for TaReSi and TaRuSi. The ratio, T$ c $/T$ _{F} $ for high T$ _{c} $ and unconventional superconductors fall in the range, 0.01$ \leq $$ \frac{T_{c}}{T_{F}} $$ \leq $0.1 \cite{umera1,umera2,umera3,umera4}. Here, this ratio $ \frac{T_{c}}{T_{F}} $ =  0.0011 and 0.0007 for TaReSi and TaRuSi places the samples in the conventional family as shown in Fig. \ref{Uemura}.

    \begin{table}[h!]
   	\caption{Superconducting and normal state parameters for TaXSi (X = Re, Ru)}
   	\label{elec propr}
   	\begin{center}
   		\begin{tabular*}{1.0\columnwidth}{l@{\extracolsep{\fill}}llll}\hline\hline
   			Parameters& unit& TaReSi& TaRuSi\\
   			\hline
   			\\[0.5ex]                                  
   			T$ _{c} $ &K &5.32 &3.91 \\
   			H$ _{c1}(0) $&mT &4.64 &6.27 \\
   			H$ _{c2} $(0)&T &1.76 &1.46 \\
   			H$ _{c2}^{P} $(0)&T &9.73  &7.15  \\
   			$\lambda_{GL} $(0)&\text{\AA} &3373 &2766 \\
   			$ \xi_{GL} $(0)&\text{\AA} &137 &114 \\
   			$ \theta_{D} $ & K & 338 & 296\\
   			$ D_{c}(E_{f}) $ & $ \frac{states}{eV f.u} $ & 2.28 & 3.34 \\
   			$ \Delta C_{el} $/$ \gamma_{n} $T$ _{c} $& & 1.07 &0.91 \\
   			$ \Delta(0) $/$ k_{B} $T$ _{c} $& &1.4 &1.36 \\
   			$m ^{*}$/m$ _{e}$ & &5.2 & 6.6 \\
   			$n$&10$^{27}$& 14 &22.2 \\
   			V$ _{\textit{f}} $&10$ ^{5} $ m/s & 1.66& 1.52 \\
   			$ {\textit{l}} $&10$ ^{-12} $m & 3.93& 2.06\\
   			$ \xi_{0} $&10$ ^{-11} $m & 3.8 & 2.42\\
   			$T_{F}$ & K & 4997& 5066
   			\\[0.5ex]
   			\hline\hline
   		\end{tabular*}
   		\par\medskip\footnotesize
   	\end{center}
   \end{table}

\section{Conclusion}
We have synthesized TaXSi (X=Re, Ru) and characterized by XRD, resistivity, magnetization, and specific heat measurements. The samples showed a type-II superconducting nature below transition temperatures 5.32 K and 3.91 K, respectively. The temperature dependence of the upper critical field determined from magnetization and resistivity has shown an upward curvature, reminiscent of a two-gap nature. A relatively high value of the H$_{c2}$(T) determined from resistivity data is likely due to surface or filamentary superconductivity. The specific heat jump around T$ _{c} $ for both the materials are less than the BCS prediction. Nonetheless, the specific heat data until 1.9 K has shown an s-wave behavior for both samples. The low value of the specific heat jump for both the materials and the upward curvature of the upper critical curve points towards a two-gap structure. However, the low value of residual resistivity and mean free path indicates strong flux pinning in this material, which in many cases can result in an upward curvature of the upper critical field. Hence, it is important to investigate this system using high quality single crystal at low temperatures to elucidate the gap structure. 

\section{Acknowledgments} R.~P.~S.\ acknowledges the Science and Engineering Research Board, Government of India for the Core Research Grant CRG/2019/001028. Financial support from DST-FIST Project No.~SR/FST/PSI-195/2014(C) is also thankfully acknowledged.

\end{document}